\documentclass[12pt]{article}
\usepackage{epsfig}

\newcommand{\beq}{\begin{equation}}
\newcommand{\eeq}{\end{equation}}
\newcommand{\beqa}{\begin{eqnarray}}
\newcommand{\eeqa}{\end{eqnarray}}
\newcommand{\eps}{\epsilon}

\begin{document}

\begin{titlepage}

\begin{flushright}
\begin{tabular}{l}
                DESY 97-029 \\
                LNF-97/007 \\
                hep-ph/9702389 \\
                February 1997   
\end{tabular}
\end{flushright}
\vspace{1.5cm}

\begin{center}
{\huge $D^*$ production from $e^+e^-$ to $ep$ collisions\\[7pt] in NLO QCD} \\

\vspace{1.5cm}

{\large Matteo Cacciari$^a$ and Mario Greco$^{b}$} \\
\vspace{.5cm}

{\sl $^a$Deutsches Elektronen-Synchrotron DESY, Hamburg, Germany\\
\vspace{.2cm}
$^b$Dipartimento di Fisica E. Amaldi, Universit\`a di Roma III,\\ and INFN, 
Laboratori Nazionali di Frascati, Italy}\\ 

\end{center}

\vspace{1cm}

\begin{abstract}
Fragmentation functions for $D$ mesons, based on the convolution of a
perturbative part, related to the heavy quark perturbative showering, and a
non-perturbative model for its hadronization into the meson, are used to 
describe $D^*$ production in $e^+e^-$ and $ep$ collisions. The non-perturbative
part is determined by fitting the $e^+e^-$ data taken by ARGUS and OPAL at 10.6
and 91.2 GeV respectively. When fitting with a non  perturbative Peterson
fragmentation function and using next-to-leading evolution for the perturbative
part, we find an $\epsilon$ parameter sensibly different from the one commonly
used, which is instead found with a leading order fit. The use of this new
value is shown to increase considerably the cross section for $D^*$ production
at HERA, suggesting a possible reconciliation between the next-to-leading order
theoretical 
predictions and the experimental data.\\[10pt]
\end{abstract}

\vfill
\noindent\rule{7cm}{.2mm}
\small
\vspace{-.4cm}
\begin{tabbing}
e-mail addresses: \= cacciari@desy.de \\
                  \> greco@lnf.infn.it 
\end{tabbing}

\end{titlepage}

\section{Introduction}

The study of fragmentation functions (FF) for heavy quarks has recently
attracted
an increased interest due to the large amount of data accumulated at LEP and
HERA. From the theoretical side predictions have been obtained by 
combining perturbative QCD - which allows to resum large
logarithms with a resulting milder renormalization/factorization scale
sensitivity - with a non perturbative component which describes the 
hadronization of the heavy quark into the meson, after the perturbative
cascade.

In $e^+e^-$ annihilation an analysis along these lines was performed by
Colangelo and Nason \cite{cn} up to LEP energies, for both charm and beauty
mesons. Due to the presence of the $c$($b$) component only, their results were
not applicable to the production of heavy mesons in hadronic collisions, where
the gluon-gluon and quark-gluon scattering play an important role. Then in a
previous analysis \cite{cgrt} a set of NLO fragmentation  functions for $D$,
$D^*$ mesons was given, including the gluon term as well, and  predictions
for large transverse momentum production cross sections were also provided.

The aim of the present analysis is to reconsider the situation of charmed meson
fragmentation functions  both in $e^+e^-$ annihilation and in photoproduction,
where new data have been obtained at HERA.

On the perturbative side, we consider the full set of perturbative
fragmentation functions (PFF's) and  their mixing in the evolution. This is
important as the OPAL data do indeed show a rise in the small $x$ region, due 
to the gluon splitting, which is  absent in the ARGUS data. In   addition, by
parametrizing the non-perturbative component by different forms and fitting
$e^+e^-$ data, we study the variation of the non perturbative parameters, in
particular for the Peterson form \cite{pssz}, as related to the accompanying
approximation, leading (LO) or next-to-leading order (NLO), used in the
perturbative component.  We find indeed that a NLO evolution favours a much
smaller value of the $\epsilon$ parameter in the non-perturbative Peterson FF
than given in the literature. In turn this also helps reconciling the recent
HERA data with the theoretical predictions. When however a LO evolution only is
considered, as in many of the parton shower Monte Carlo codes used in the
experimental analyses, the "conventional" value for $\epsilon$ is recovered.
This result can be understood by noting that the effect of parton showering,
which is larger in a NLO analysis, softens the distribution of the partons, 
acting qualitatively as a non perturbative FF, which can henceforth behave more
softly. Therefore the value of $\epsilon$ used in the phenomenological analyses
must be closely related to the level of the approximation followed in the
perturbative QCD evolution.

This paper is structured as follows: in Section \ref{theory} we recall the
theoretical framework, partly already introduced in \cite{cgrt}, on which this
work is based. Section \ref{ee} presents the results of fits to ARGUS and OPAL
data in $e^+e^-$ collisions. Section \ref{ep} makes use of the non perturbative
parameters previously determined to give predictions for $D^*$ photoproduction
in $ep$ collisions at HERA. Our conclusions are then given in Section
\ref{concl}.

\section{Theoretical Framework}
\label{theory}

We have already introduced in Ref. \cite{cgrt} the theoretical framework for
evaluating $D$ mesons cross section within a fragmentation approach. In that
paper, the following ansatz for the fragmentation function (FF) of a parton 
$i$ into a meson $D$ was made:
\beq
D_i^D(x,\mu) = D_i^c(x,\mu) \otimes D_{np}^D(x).
\label{ansatz}
\eeq
In this equation, $D_i^c(x,\mu)$ is the perturbative fragmentation function
(PFF) for a massless parton to fragment, via a perturbative QCD cascade, into 
the massive charm quark $c$. $D_{np}^D(x)$ is instead a non-perturbative
fragmentation function, describing the transition from the heavy quark to the
meson. Finally, the symbol $\otimes$ indicates convolution, i.e.
\beq
f(x) \otimes g(x) \equiv \int_x^1 {{dz}\over{z}} f(z) g(x/z).
\eeq

The formalism of PFF's has been introduced a few years ago \cite{melenason},
and will not be given here in detail. We just recall that it allows
to extract from perturbative QCD (pQCD)  the initial state conditions for the
PFF's at a scale $\mu_0$ of the order of the heavy quark mass
$m$ (and we will take $\mu_0 = m$):
\begin{eqnarray}
&&D_c^c(x,\mu_0) = \delta(1-x) + {{\alpha_s(\mu_0) C_F}\over{2\pi}}\left[
{{1+x^2}\over{1-x}}\left(\log{{\mu_0^2}\over{m^2}} -2\log(1-x)
-1\right)\right]_+ \label{DQQ} \\ 
&&D_g^c(x,\mu_0) = {{\alpha_s(\mu_0) T_F}\over{2\pi}}(x^2 + (1-x)^2)
\log{{\mu_0^2}\over{m^2}} \label{DgQ} \\
&&D_{q,\bar q,\bar c}^c(x,\mu_0) = 0 \label{DqQ}
\end{eqnarray}
where $c$ represents here the heavy quark and $g$ and $q$ the gluon and light
quarks respectively. Moreover, $C_F = 4/3$ and $T_F = 1/2$.

The PFF's, evolved up to any scale $\mu$ via the
Dokshitzer-Gribov-Lipatov-Altarelli-Parisi (DGLAP) equations, 
can be used to evaluate heavy quark cross sections in the
large transverse momentum ($p_T$) region (i.e. $p_T \gg m$) by convoluting them
with cross section kernels for massless partons \cite{aversa, aurenche,
aurenche2}, subtracted in the modified minimal subtraction ($\overline{MS}$)
scheme, where the heavy quark is also treated as a massless active flavour and
therefore also appears in the parton distribution functions of the colliding
hadrons.
This has been done in Ref. \cite{cg1} for $p\bar p$, in Ref. \cite{cg2} for
$\gamma p$ and finally in Ref. \cite{cgkkks} for $\gamma\gamma$ collisions. In
all cases it has been shown how the results agree with the full massive ones
(Refs. \cite{NDE}, \cite{EN} and \cite{michael} respectively) in an
intermediate $p_T$ region (say from twice to four times the mass of the heavy
 quark). For larger $p_T$ they are more reliable (and hence have a
smaller scale dependence) because the large logarithms originating from
gluon emission and gluon splitting are resummed by the evolution of the PFF's 
(see Ref. \cite{cg1} for a more complete discussion on this point).

The fragmentation functions of eq. (\ref{ansatz}) will be also evolved with the
DGLAP equations. It is to be noted
that in doing so we assume the evolution to be entirely perturbative in
character: we evolve the full FF's (\ref{ansatz}) as we would the PFF's only.
The non-perturbative part of the overall FF's is kept fixed and 
determined at a given experiment.

Indeed, the non-perturbative part of the FF's cannot  be
predicted by pQCD. In fact, the process through which a heavy quark binds to a
light one to form the meson involves exchanges of gluons with momenta of order
$\Lambda_{QCD}$ or smaller, and is therefore intrinsically non-perturbative.
However, a few features of this function can be determined. In contrast to
light quark hadronization, this FF is hard \cite{hard} because the meson
retains a larger fraction of the heavy quark initial momentum. Moreover, one
expects the non-perturbative FF to be squeezed towards $x=1$ linearly in the
mass of the heavy quark. This statement is proved in \cite{nason} under the
hypothesis of softness of the hadronization process and in the infinite mass
limit (see also \cite{kz} for a discussion on this point).

In the following we will employ two different functional forms for the
non-perturbative part of the fragmentation function.

The first one is dictated mainly by its semplicity, and is given by
\beq
D_{np}(x;\alpha,\beta) = A (1-x)^\alpha x^\beta
\label{simple}
\eeq
with
\beq
{1\over A} = \int_0^1 (1-x)^\alpha x^\beta dx = B(\beta+1,\alpha+1),
\eeq
$B(x,y)$ being the Euler Beta function. This functional form had already been
employed in \cite{cn} for fits to $e^+e^-$ data and was also used in our
previous paper on charmed meson FF's \cite{cgrt}. It is flexible enough to
describe the data and has the advantage of an easily calculable Mellin
transform, given by
\beqa
D_{np}(N;\alpha,\beta) &\equiv& \int_0^1 dx x^{N-1} D_{np}(x;\alpha,\beta) =
\nonumber\\
&=&{{B(\beta+N,\alpha+1)}\over{B(\beta+1,\alpha+1)}} = {{\Gamma(\beta+N)
\Gamma(\alpha+\beta+2)}\over{\Gamma(\beta+1)\Gamma(\alpha+\beta+N+1)}},
\eeqa
with $\Gamma(x)$ being the Euler Gamma function.

However, this functional form has no immediate physical motivation. A
successful description of $e^+e^-$ data could be not enough to ensure the
correctness of the predicted cross sections in, say, $ep$ production 
evaluated with the same
non-perturbative FF, since higher moments could play an important role. Indeed,
in $e^+e^-$ collisions it is the mean scaled energy, i.e. $\int dz z D(z)$ - or
the second moment when talking Mellin transforms language - the most important
observable. Different FF's could therefore agree on this second moment but
then have different higher moments which could lead to different prediction in
other kinds of reactions.

We have therefore chosen to employ also a different non-perturbative
fragmentation, based on a physical model: the so called
Peterson form \cite{pssz}. It is derived by considering the transition
amplitude for a fast moving heavy quark $Q$ to fragment into $(Q\bar q) + q$,
$q$ being a light quark. It reads
\beq
D_{np}(x;\eps) = {A\over{x\left[1-1/x-\eps/(1-x)\right]^2}},
\label{peterson}
\eeq
with the normalization factor $A$ now given by
\beq
{1\over A} =
\frac{(\epsilon^2-6\epsilon+4)}{(4-\epsilon) \sqrt{4\epsilon-\epsilon^2}}
\left\{
\arctan\frac{\epsilon}{\sqrt{4\epsilon-\epsilon^2}}
+ \arctan\frac{2-\epsilon}{\sqrt{4\epsilon-\epsilon^2}} \right\}
+ \frac{1}{2} \ln \epsilon + \frac{1}{4-\epsilon}.
\eeq
From the derivation one finds that the $\eps$ parameter is related to the heavy
quark mass by $\eps \simeq \Lambda^2/m^2$, where $\Lambda$ stands for a
hadronic scale. Since the average scaled energy goes like $\langle x\rangle =
1-\sqrt{\eps}$, we see it respects the prediction of scaling linearly with the
heavy quark mass.

While this form of non-perturbative fragmentation function is certainly more
physical and the order of magnitude of its unknown parameter can be estimated
from first principles, it has however the drawback of a much more complicated
Mellin transform. The full expression is given in the Appendix of Ref.
\cite{kks}, and will not be repeated here.

\section{Production in $e^+e^-$ collisions}
\label{ee}

According to QCD factorization theorems, the cross section for the production
of a hadron H in the $e^+e^-$ process
\beq
e^+e^- \to \gamma,Z\to H\,X,
\eeq
at a center-of-mass energy $Q = \sqrt{s}$, can be written as
\beq
{{d\sigma^H}\over{dx}} = \sum_i\int_x^1 {dz\over z}C_i(z,\alpha_s(\mu),Q,\mu)
D_i^H\left({x\over z},\mu\right) \equiv \sum_i C_i(z,\alpha_s(\mu),Q,\mu)
\otimes D_i^H(z,\mu),
\label{fact}
\eeq
$x$ being the energy fraction of the produced hadron, $x=2E/Q$. The functions 
$C_i(z,\alpha_s(\mu),Q,\mu)$ are the so called coefficient functions, which
describe the hard part of the scattering process and can be calculated in
perturbation theory as series expansions in the strong coupling $\alpha_s(\mu)$.
Explicit expressions up to NLO for all the coefficient functions we need can 
be found, for instance, in Ref. \cite{nw}.
Since we take the partons in the hard scattering to be massless, collinear
singularities appear, and these are subtracted in the $\overline{MS}$ scheme
and reabsorbed into the fragmentation functions. $\mu$ is the
factorization scale at which this subtraction is performed, which in this case
we have for simplicity taken equal to the renormalization scale. The sum is to
run on all the partons which can be considered massless in the coefficient
functions. Since in general mass terms of the form of powers of $m/Q$ will
appear, we see that already at $Q=10$ GeV the charm can to a good approximation
be taken as massless. The same will be true also for the bottom quark at $Q=91$
GeV, whereas its production should instead be strongly suppressed at the lower
energy. We will therefore include four and five active flavours respectively at
these two center-of-mass energies.

When dealing with light hadrons the fragmentation functions can only be
determined by comparison with experiment. Since in our case the hadron in
question is instead the heavy meson $D^*$, we can make use of our ansatz of eq.
(\ref{ansatz}), and fit to the experimental data only the non-perturbative part
of the FF's.

We start by trying to fit the non-perturbative FF to experimental data for
$D^{*\pm}$ production taken by ARGUS \cite{argus} and OPAL \cite{opal} at 10.6
GeV and 91.2 GeV respectively. The cross section is evaluated by means of the
formula in eq.(\ref{fact}), the fragmentation functions are given by the
initial conditions reported in the previous section, evolved up to the
desired 
scale with the DGLAP equations to next-to-leading (NLO) order and convoluted
with the non-perturbative component.
 
\subsection{Fits with $(1-x)^\alpha x^\beta$} 
 
We first perform fits with the ``simple'' form $(1-x)^\alpha x^\beta$. Similar
fits had already been performed a few years ago in Ref. \cite{cn}. In that
paper only the non-singlet component of the FF's had been taken into account, a
valid approximation at the low energy of 10.6 GeV. When going to higher energy,
on the other hand, the mixing with the gluons through the evolution will become
more and more important. The OPAL data do indeed show a rise in the small $x$
region, due to gluon splitting and absent in the ARGUS data. We have therefore 
included the full set of FF's and mixings in the evolution.

\begin{table}
\begin{center}
\begin{tabular}{|l c c c|}
\hline
         & $\alpha$ & $\beta$  & $\chi^2$/d.o.f\\
\hline
\hline
\multicolumn{4}{|c|}{With Sudakov resummation}\\
ARGUS, Ref. \cite{cn} & 0.4              &  4.6             &  \\
ARGUS                 & 0.51 $\pm$ 0.37  &  4.9 $\pm$ 1.7   &  0.70 \\
OPAL                  & 0.30 $\pm$ 0.21  &  4.5 $\pm$ 1.5   &  1.26\\
\hline
\multicolumn{4}{|c|}{Without Sudakov resummation}\\
ARGUS        & 1.0 $\pm$ 0.6    & 6.7 $\pm$ 2.3   &  0.86    \\
OPAL         & 0.9 $\pm$ 0.3    & 6.4 $\pm$ 1.9   &  1.32 \\
\hline
\end{tabular}
\parbox{13cm}{
\caption{\label{table0}\small Results for the fitting of $\alpha$ and $\beta$
in $(1-x)^\alpha x^\beta$ to ARGUS and OPAL data. Evolution is performed 
to NLO and with $\Lambda_5$ = 200 MeV and $\mu_0=m$.}
}
\end{center}
\end{table}

As a first step, we have refitted the same ARGUS data already considered in
Ref. \cite{cn}. We have taken $\Lambda_5$ = 200 Mev and included in the PFF's
the resummation of Sudakov terms in the $x\simeq 1$ region, as described in
\cite{melenason} and consistently with \cite{cn}. A normalization factor is
always fitted along with the parameters determining the shape of the
non-perturbative FF. The results are shown in the upper part of Table
\ref{table0}.

They can be seen to be consistent with those obtained in Ref. \cite{cn}.
It is also worth mentioning that the last point in the ARGUS data has not been
included in our fit. In that region non-perturbative effects become very large,
spoiling the evaluation of the perturbative part of the FF's: the PFF's evolved
to NLO become indeed negative in the large $x$ region.
We have therefore preferred not to include that point in the fit.

We have also presented along with the fits to ARGUS the results of a similar
fit to OPAL data. Also in this case a few points have been excluded from the
fit: the last one, where again large non-perturbative effects set it, and the
first three ones, where the rise due to gluon splitting is observed. Since
unaccounted for threshold effects may play an important role here, and the
theoretical curve cannot be made to describe the data very well, we have
preferred to avoid biasing the fitted parameters and therefore excluded this
region altogether.

The main result is the consistency of the two sets of parameters: the same
values which fit the ARGUS data also describe the OPAL data, taken at a
center-of-mass energy almost one order of magnitude larger. This finding lends
support to our initial hypothesis of scale independence of the non-perturbative
part of the fragmentation functions.

\begin{figure}[t]
\begin{center}
\epsfig{file=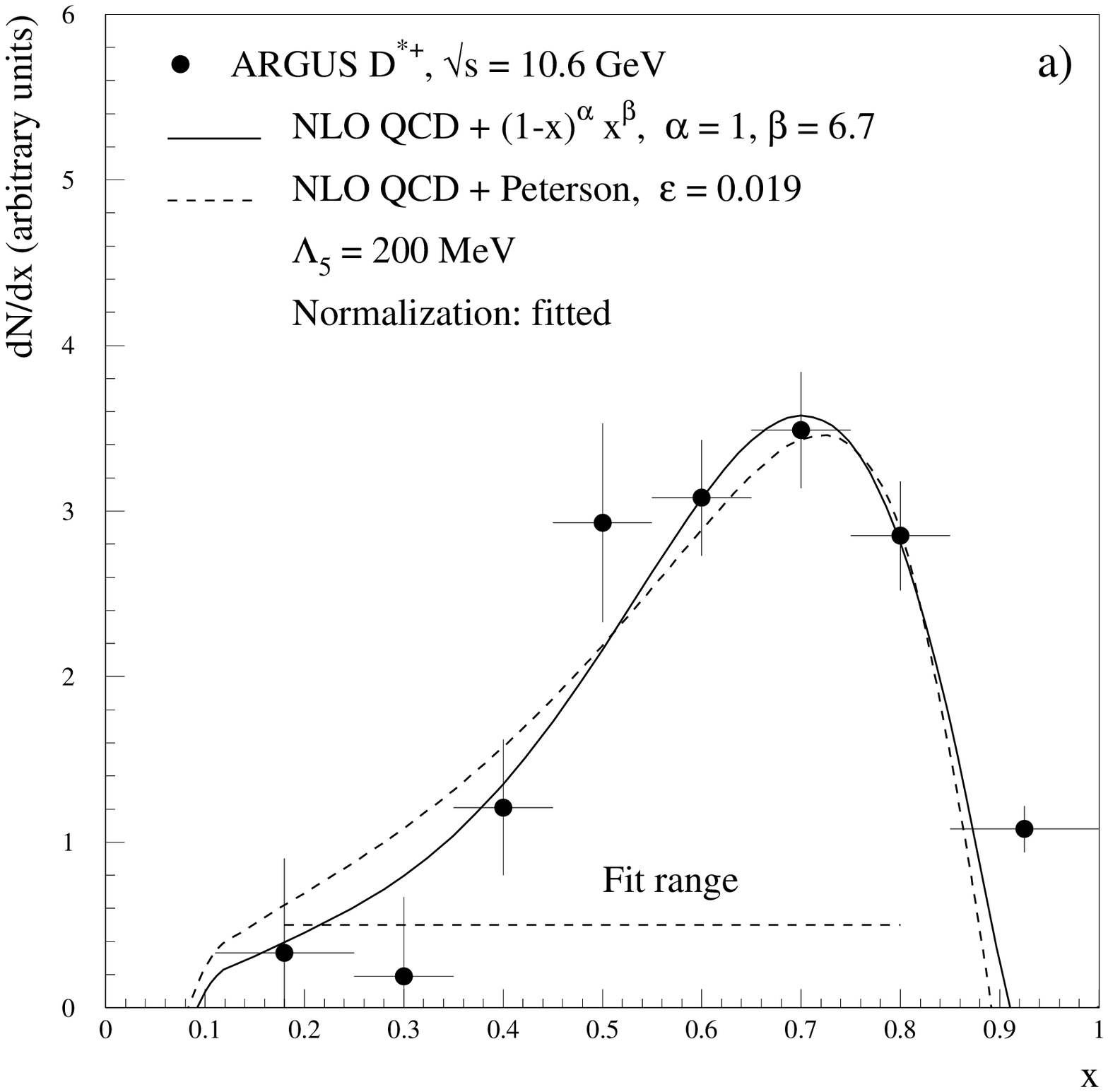,
              bbllx=30pt,bblly=160pt,bburx=540pt,bbury=660pt,
             width=9.5cm,clip=}
\hspace{.5cm}
\epsfig{file=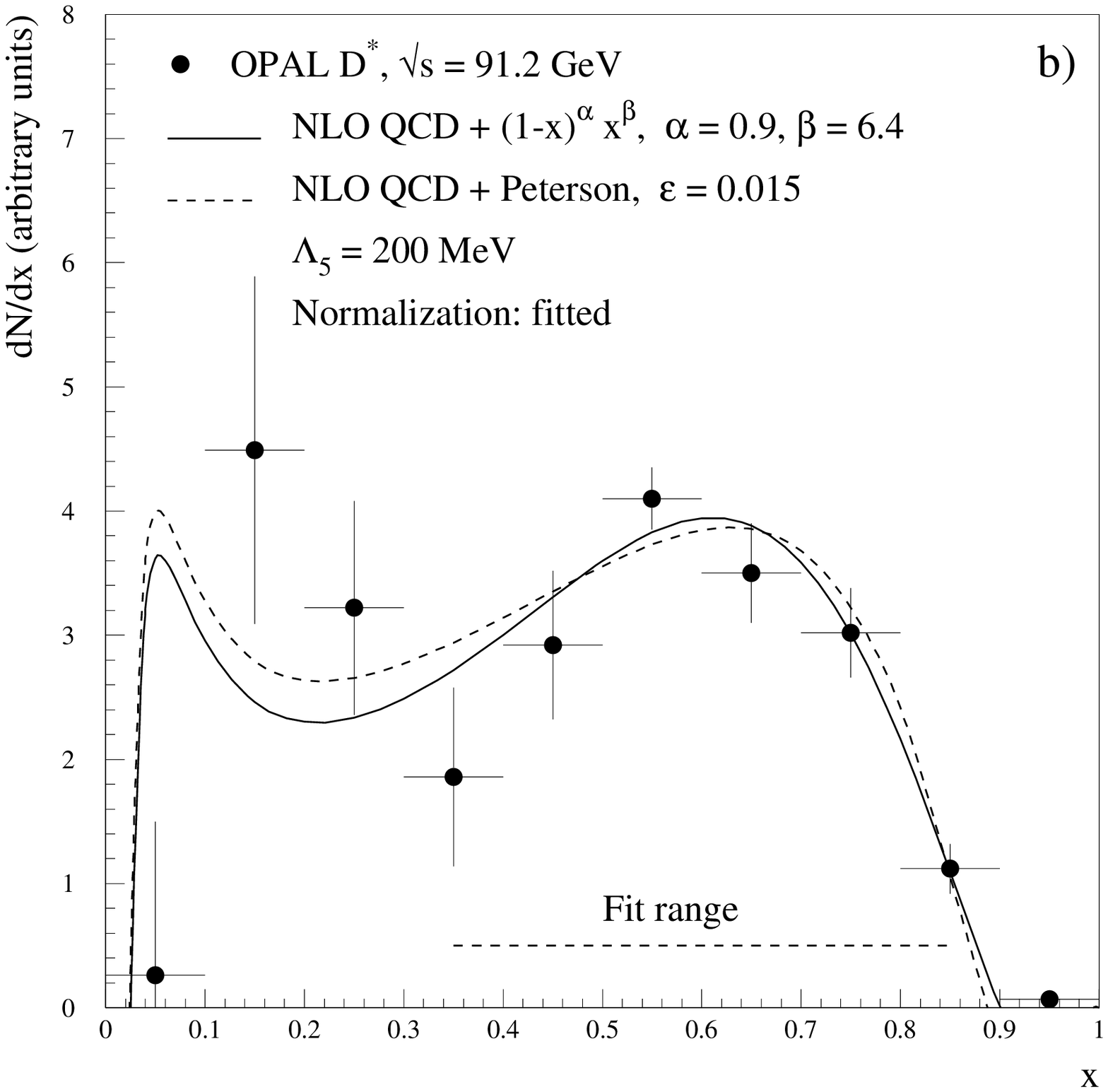,
             bbllx=30pt,bblly=160pt,bburx=540pt,bbury=660pt,
             width=9.5cm,clip=}
\parbox{13cm}{
\caption{\label{fig1}\small Distributions of $D^*$ mesons as measured by the
ARGUS and OPAL experiments, together with the theoretical curves fitted to 
the same data with
the $(1-x)^\alpha x^\beta$ (full line) and the Peterson (dashed line)
non-perturbative fragmentation functions.
}
}
\end{center}
\end{figure}

Other fits with this ``simple'' non-perturbative FF have been performed, this
time excluding the resummation of Sudakov terms. The reason for this is that
when making convolutions of the PFF's with the Sudakov included in the $x$
space (rather than in Mellin moments space as we do now) the integration
convergence is much more difficult. We have therefore chosen to incorporate the
effect of the Sudakov resummation into the non-perturbative part, with the
results given in the lower part of Table \ref{table0}. Once more,
full consistency is found between the fits to ARGUS and to OPAL data. The
results of these two fits are shown in figure \ref{fig1}.

\subsection{Fits with the Peterson form} 

Fits to the same ARGUS and OPAL data have also been performed using the
Peterson form (\ref{peterson}) as the non-perturbative part of the FF's. The
fit is in this case a two- rather than a three-parameter one, namely the
normalization and the $\eps$ parameter only. Using NLO evolution and
coefficient functions, but again no Sudakov resummation, and three 
different values for $\Lambda_5$, we have found the results displayed in Table
\ref{table1}, while the  curves resulting from these fits, for the choice 
$\Lambda_5$ = 200 MeV, are shown in figure \ref{fig1}.

It is to be noted that the fitter was not able, in a few instances, to produce
realistic errors when fitting ARGUS data, due to  numerical
inaccuracies resulting from the inverse Mellin transform of the Peterson FF.
However, taking the error in the corresponding fit to OPAL data as an
indication, we see that also in this case the two fits are consistent, pointing
to a scale independence of the non-perturbative part of the fragmentation
functions. 

The most striking feature of these fits is however the discrepancy between
their results and the value commonly used for the parameter $\epsilon$ when
describing $c$ quarks fragmentation to $D^*$ mesons. It is indeed found in the
literature (see, for instance, Ref. \cite{chrin}), and has been used in recent
phenomenological papers \cite{fnr,kks}, the value $\eps = 0.06$. The fitted
values (except for the one at $\Lambda_5$ = 100 MeV) also appear  to be
at variance with the result found by the OPAL Collaboration 
\cite{opal} as a fit to their own data,
$\eps_\mathrm{OPAL} = 0.035 \pm 0.007 \pm 0.006$.

\begin{table}
\begin{center}
\begin{tabular}{|l c c c|}
\hline
     &$\Lambda_5$ = 100 MeV&$\Lambda_5$ = 200 MeV&$\Lambda_5$ = 300 MeV\\
\hline
\hline
\multicolumn{4}{|c|}{Next-to-leading order evolution}\\
ARGUS  & .031 (1.09) &.019 (1.27)     & .011 $\pm$ .003 (1.53)\\
OPAL   &.033 $\pm$ .005 (1.25) &.015 $\pm$ .002 (1.54) &.008 $\pm$ .001
(1.72)\\
\hline
\multicolumn{4}{|c|}{Leading order evolution}\\
ARGUS  & .07 (1.65) & .055 (2.1) & .036 (2.72) \\
OPAL   & .10 $\pm$ .01 (2.02)  &  .08 $\pm$ .01 (2.48) &.06 $\pm$ .01 (2.98)\\
\hline
\end{tabular}
\parbox{13cm}{
\caption{\label{table1}\small Results for the fitting of the $\eps$ parameter
of the Peterson FF to ARGUS and OPAL data, for three different values of
$\Lambda_5$ and with next-to leading order coefficient functions and NLO or LO 
evolution of the PFF's.
Sudakov resummation is not included explicitly, and is therefore effectly
reabsorbed into the non-perturbative FF. The number between the round brackets
is the $\chi^2$ per degree of freedom of each fit.}
}
\end{center}
\end{table}

This discrepancy should however not come as a surprise if one considers
carefully how $\eps$ so far has been extracted from experimental data. 
Experiments usually report the energy or momentum fraction ($x_E$ or $x_p$) of
the observed hadron with respect to the beam energy. On the other hand the
momentum fraction which appears as the argument of the non-perturbative FF is
rather the fraction with respect to the fragmenting quark momentum, usually
denoted by $z$ (see for instance \cite{chrin} for a discussion on this point).
These two fractions are not coincident, due to radiation processes  which lower
the energy of the quark before it fragments into the hadron. In order to
deconvolute these effects one usually runs a Monte Carlo simulation of the
collision process at hand, including both the parton showers and the subsequent
hadronization of the partons into the observable hadrons. The latter can be
parametrized in the Monte Carlo by the same Peterson fragmentation function we
have been using, and the value of $\eps$ which best describes the data can be
extracted. But what can be different in our approach is of course the
perturbative QCD part, namely the parton shower. This showering softens the
distribution of the partons, producing an effect qualitatively similar to that
of the non-perturbative FF. On the quantitative level, the amount of softening
(and hence the value of $\eps$) required by the non-perturbative FF to
describe  the data is related to the amount of softening already performed at
the  perturbative level. Monte Carlo's simulations so far only perform a
leading order description of the showering, and can hence differ  from our NLO
evolution.

Therefore there is not   a ``unique'' and ``true'' value for the parameter
$\eps$, but only a value closely interconnected with the details of the
description of the pQCD part of the problem. For instance, a higher value of
$\Lambda_5$ results in a larger $\alpha_s$ and hence in more parton showering.
This softens even more the perturbative part of the FF, and consequently less
softening will be required from the non-perturbative part. The results in Table
\ref{table1} show that this is indeed the case, a smaller value of $\eps$
corresponding to a harder Peterson FF.

A double check that the different description of the perturbative part can
indeed responsible for the different $\eps$ can be done by rerunning our fits
with a leading order evolution, in such a way to mimick as closely as possible
the Monte Carlo description of the process. The results are displayed in Table
\ref{table1}, and can be seen to be indeed much closer to the commonly used
value of 0.06. The tendency to a discrepancy between ARGUS and OPAL fits could
actually be an indication of the inadequacy of a leading order description of
the scale violations taking place from 10 to 90 GeV. All this should however
not be taken literally, as many other details might be included in the leading
order Monte Carlo description of the perturbative showering and be missing or
differently treated here.

A further check of the modification of the  value for $\eps$ when going
from a leading to a next-to-leading description of the perturbative parton
shower can be obtained in the following way, to be taken as a kind of
toy-model.

Consider a distribution for the energy variable $x$, like the ones given by
ARGUS and OPAL and plotted in fig. \ref{fig1}. Thinking of them as described by
the
convolution of a perturbative and a non perturbative fragmentation function,
the average value of $x$, call it $\langle x \rangle_{exp}$, can be written as a
product of the average values of the perturbative and the non-perturbative
FF's, i.e.
\beq
\langle x \rangle_{exp} = \langle x \rangle_{pert}\langle x \rangle_{np}.
\eeq
If we now assume that both a leading and a next-to-leading description of the 
perturbative part can describe the data, provided they are matched by the
appropriate non-perturbative FF (i.e., the appropriate value of $\eps$ is
chosen), we can write
\beq
\langle x \rangle_{exp} = \langle x \rangle_{pert}^{LO}\langle x
\rangle_{np}^{LO} = 
\langle x \rangle_{pert}^{NLO}\langle x \rangle_{np}^{NLO},
\eeq
which leads us to
\beq
\langle x \rangle_{np}^{NLO} = {{\langle x \rangle_{pert}^{LO}}\over
{\langle x \rangle_{pert}^{NLO}}} \langle x\rangle_{np}^{LO}.
\label{xnlo}
\eeq
In this equation $\langle x \rangle_{pert}$ refers to the second Mellin
moment of the perturbative fragmentation function $D_c^c$, while the
$\langle x\rangle_{np}$ can be calculated from the Peterson FF, like $\langle
x\rangle_{np} = \int x
D(x;\eps) dx$. The suffixes ``LO'' and ``NLO'' on the perturbative parts mean
that a leading or next-to-leading evolution kernel has been included before
taking the average. The non-perturbative part is considered to
be adjusted to fit the data together with the given perturbative term.

The perturbative fragmentation function returns the following averages when
evolved with $\Lambda_5 = 200$ MeV:
\begin{center}
\begin{tabular}{l c c c}
       & $\langle x \rangle^{LO}_{pert}$ & $\langle x \rangle^{NLO}_{pert}$ &
  $\langle x \rangle_{pert}^{LO}/\langle x \rangle_{pert}^{NLO}$\\[5pt]
10.6 GeV           & .75  & .65  &  1.15 \\
91.2 GeV           & .64  & .56  &  1.14
\end{tabular}
\end{center}
We can clearly see from this table how the NLO description does indeed soften
the perturbative FF more than the LO one,  producing a lower value for 
the average energy.

Assuming $\eps=0.06$ to be the right value to describe the data when a leading
order perturbative description is used, we get $\langle x\rangle_{np}^{LO} =
0.67$ and hence, from eq. (\ref{xnlo}), $\langle x\rangle_{np}^{NLO} = 0.77$.
Upon inspection we see this average value for the Peterson FF corresponds
to $\eps = 0.016$, i.e. a value fully compatible with the ones returned by the
fits.

Before closing this Section on the fits, we wish to point out once more that
there is not a ``best candidate'' value for  $\eps$, but only a value of $\eps$
more suited to match the description of the perturbative showering one is
actually employing. Surely enough, if the QCD description is at NLO a
harder $\eps$, like our $\eps=0.015$,  should be used
rather than the larger (and softer) $\eps = 0.06$, since part of
the softening is now already included through more perturbative gluon emission.


\section{Production in $ep$ collisions}
\label{ep}

The use of fragmentation functions for heavy quarks to evaluate NLO cross
sections for charm photoproduction has already been considered in Ref.
\cite{cg2}.

In this paper we use exactly the same formalism to evaluate cross sections for
$D^*$ production, by complementing the PFF's used in the previous work with a
non-perturbative component as described by eq. (\ref{ansatz}) and according to
Ref. \cite{cgrt}.  

The $\gamma p$ cross section reads, schematically,
\beq
d\sigma_{\gamma p} = \int F_{i/p} d\hat\sigma_{\gamma i\to k} D_k^D +
                 \int F_{i/p} F_{j/\gamma} 
                 d\hat\sigma_{ij\to k} D_k^D.
\eeq
In this expression $F_{i/p}$ and $F_{j/\gamma}$ are the  parton distribution
functions (pdf's) for the proton and the photon, since the so called direct and
resolved component are both included. Unless otherwise stated, we will make use
of the MRS-G \cite{mrsg} and ACFGP \cite{acfgp} sets respectively. The
$\hat\sigma$'s are the kernel cross sections (= coefficient functions)  for
massless parton production \cite{aversa, aurenche} and $D_k^D$ is the meson
fragmentation function of eq. (\ref{ansatz}). We will use in this FF the
non-perturbative parameters fitted in the previous section to $e^+e^-$ data
and, since the non-perturbative FF's are normalized to one, we include the
branching ratio $BR(c\to D^{*+}) = BR(\bar c\to D^{*-}) = 0.26$ \cite{opal}.
This produces an absolute, parameter free, prediction, to be directly compared
with  the experimental data.

We also convolute our $\gamma p$ cross sections with the 
Weizs\"acker-Williams flux factor,
\beq
\sigma_{ep}(s) = \int_{y_{min}}^{y_{max}} dy f_{\gamma/e}(y) 
\sigma_{\gamma p}(ys)
\eeq
with
\beq
f_{\gamma/e}(y) = {\alpha\over{2\pi}} \left[ {{1 + (1-y)^2}\over{y}}
\ln{{Q_{max}^2}\over{Q_{min}^2}} + 2 m_e^2 y \left( {1\over{Q_{max}^2}}
-{1\over {Q_{min}^2}}\right)\right],
\eeq
where $y=E_\gamma/E_e$, $Q_{min}^2 = m_e^2 y/(1-y)$ and $m_e$ is the 
electron mass, to mimick as closely as possible
the experimental setup. For comparisons with ZEUS data we will
adopt $Q_{max}^2 = 4$ GeV$^2$  and $y_{min} = 0.147$, 
$y_{max} = 0.869$, according to \cite{zeuswwa}. 
Moreover, we will present cross sections in the pseudorapidity range $-1.5
<\eta<1$ and in the $p_T$ range $3<p_T<12$ GeV. 

\begin{figure}[t]
\begin{center}
\epsfig{file=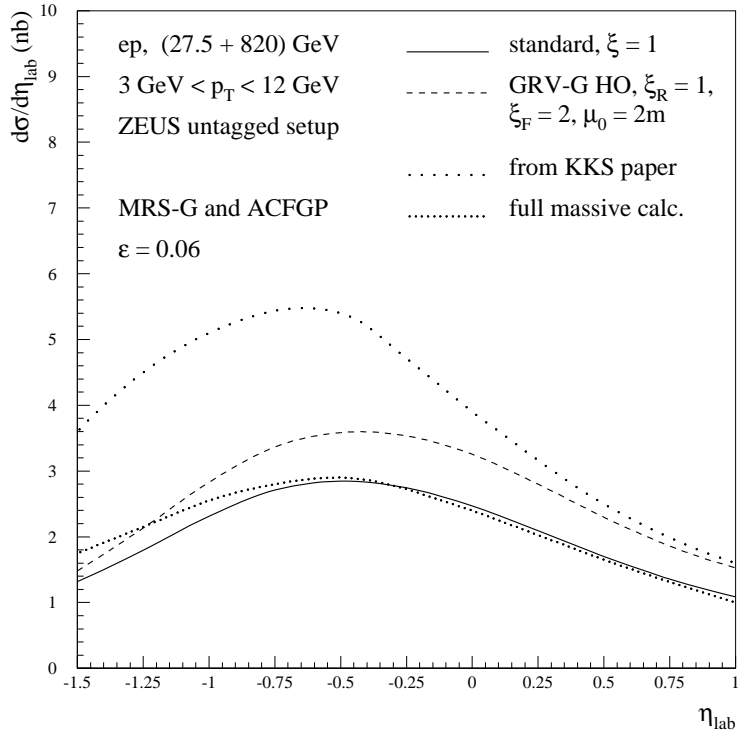,
              bbllx=30pt,bblly=160pt,bburx=540pt,bbury=660pt,
             width=10cm,clip=}
\parbox{13cm}{
\caption{\label{fig4}\small Comparison of our results with those of Ref.
\protect\cite{kks} (KKS) and with the full NLO massive calculation of Ref.
\protect\cite{fnr}. The GRV-G HO photon parton 
distribution functions set is employed for all the curves except for the 
``standard'' one (full line). $\xi_R$ and $\xi_F$ refer to the ratios of
the renormalization and factorization scales to the transverse 
mass $m_T$ respectively.
}
}
\end{center}
\end{figure}

As already stressed in Ref. \cite{cgrt},
it is important to point out how this low $p_T$ boundary casts doubts on the
validity of an approach based on the use of massless cross section kernels, and
which had originally been devised for the resummation of large logarithms in
the large $p_T$ region. In principle, one is missing terms of order
$m/p_T$, and the errors may therefore be large when $p_T \simeq m$. Only a
comparison with a full massive calculation can finally assess whether the
results are meaningful enough. Such a comparison will be presented 
in fig. \ref{fig4}.

A description of $D^*$ photoproduction in $ep$ collisions similar to ours has
recently been given in \cite{kks}. When including the Peterson FF these authors 
tackle
the problem from an apparently different point of view, by evolving directly
this non-perturbative FF and inserting instead the initial conditions
(\ref{DQQ}), (\ref{DgQ}) and (\ref{DqQ}) for the heavy quark PFF's into the
coefficient functions for $\gamma p$ to massless parton scattering. One can
however easily see the two approaches are equivalent at the perturbative
level. The Appendix does indeed show how they should only differ
by uncontrollable higher order terms and, other than this, in the
interpretation of the various components.

Therefore the approach introduced in Ref. \cite{cgrt} and now discussed here in
detail,  and the one successively used in \cite{kks} should give similar
results.  We compare them in fig. \ref{fig4}.  It shows the curve
extracted from Ref. \cite{kks} (wide-dotted line) and our results, for the same
value of $\eps=0.06$. No agreement is found, however, neither (full line) with
what will be our standard choice of renormalization/factorization scales ($\mu
= \mu_R = \mu_F = \xi m_T = \xi \sqrt{m^2 + p_T^2}$ with $\xi=1$, $\mu_0=m$,
and $\Lambda_5$ = 200 MeV) nor (dashed line) when we make the same choice as
Ref. \cite{kks}, taking $\mu_R = m_T$, $\mu_F = 2m_T$, $\mu_0=2m$, GRV-G HO 
\cite{grvg} as the photon pdf's set.

Spurious higher order terms  could be responsible for the discrepancy. If
one does indeed check fig. 2 of Ref. \cite{kks}, by comparing curves C and D a
difference similar to the one found above between the wide-dotted and the
dashed line can be seen. This large difference could therefore be due to the
moving of the initial condition terms for the fragmentation function to the
kernel cross sections for massless parton scattering (see Appendix). 
Curve D of Ref.
\cite{kks} has been made following our standard PFF formalism, 
and by comparing it with our results we have indeed found agreement.

It is worth noting that the spurious terms contain large Sudakov logarithms of
the form  $\log(1-x)$, and could indeed be not negligible. Since we fitted
$e^+e^-$ data with the same overall fragmentation function we are now using
here, we believe the large effect of these terms - if present - to be
effectively absorbed  into the fixed non-perturbative component. Hence it
should not spoil a reliable evaluation of photoproduction cross sections.

Also shown on the same plot is the result of the full NLO massive calculation
by \cite{fnr} (close-dotted line), itself convoluted with a Peterson FF with
$\eps=0.06$ too, as taken from \cite{zeuswwa}. Good agreement with our result 
is found, especially when making our standard choice of scales. Such a
successful comparison could probably not have been expected beforehand,  given
the missing $m/p_T$ terms, but a posteriori it can perhaps be considered a
check of our results, being the massive result the benchmark at these low $p_T$
values.  The agreement will also allow us to extrapolate to the massive
calculation the effect of varying the  value of $\eps$.

\subsection{Comparison with experiment}

\begin{figure}[t]
\begin{center}
\epsfig{file=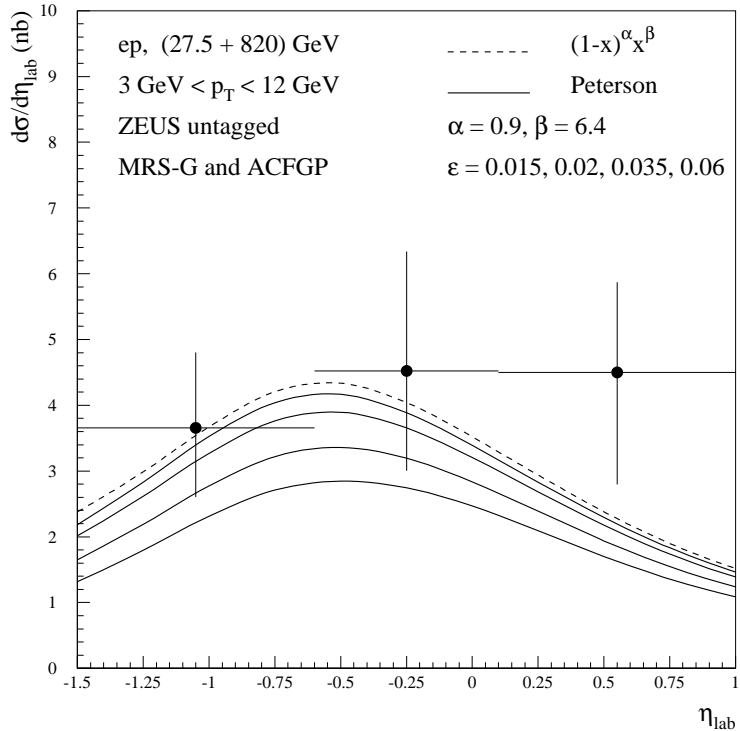,
              bbllx=30pt,bblly=160pt,bburx=540pt,bbury=660pt,
             width=10cm,clip=}
\parbox{13cm}{
\caption{\label{fig2}\small Pseudorapidity distribution of $D^*$ as measured by
the ZEUS experiment and theoretical predictions for
different values of $\eps$ in the Peterson fragmentation function (full line)
and with the $(1-x)^\alpha x^\beta$ FF (dashed line).
}
}
\end{center}
\end{figure}

We now compare our results with experimental photoproduction data obtained at
HERA by ZEUS \cite{zeuswwa} and H1 \cite{h1} Collaborations.

We first plot, in fig. \ref{fig2}, the pseudorapidity distributions, integrated
over the $p_T$, obtained with the Peterson FF with different values of $\eps$.
These results have been obtained with the pdf's set
MRS-G for the proton, and ACFGP for the photon in the resolved component. For
the renormalization/factorization scales we made the standard choice $\mu =
\mu_R = \mu_F = m_T$ and taken $\mu_0=m$ as the starting value for the
evolution of the FF's. $\Lambda_5$ is taken equal to 200 MeV.

As expected, the use of a smaller $\eps$ hardens the non-perturbative FF and
hence enhances the cross section, since the partonic kernels fall rapidly with
increasing $p_T$. The cross section obtained with $\eps=0.015$ is 50\% larger
than that with $\eps=0.06$, and while the latter seems to fall short of
describing the ZEUS data, the former does a good job, at least in the first two
bins. But we emphasize here once more how a full assessment of the reliability
of these results needs a comparison with the full massive calculation, rather
than with the experimental data, which however need to be improved in precision.

For comparison, the cross section obtained with the simple FF, $(1-x)^\alpha
x^\beta$, with $\alpha = 0.9$ and $\beta=6.4$, is also shown (dashed line) in
fig. \ref{fig2}.  These values for $\alpha$ and $\beta$ fit the OPAL data from
$e^+e^-$ collisions like $\eps=0.015$ does, see Section \ref{ee}, and the
photoproduction cross sections are indeed also in good agreement. This on one
side shows how in this case there is little dependency on the precise shape of
the non-perturbative fragmentation function. On the other side, it strengthens
our trust of the cross section with the Peterson, much harder to evaluate 
due to the numerical difficulties related to the inverse Mellin transform.

\begin{table}
\begin{center}
\begin{tabular}{|l|c|}
\hline
                        &  $\sigma$ (nb) \\
\hline
$\alpha=0.9,~\beta=6.4$ &    8.0\\
$\eps = 0.015$          &    7.7 \\
$\eps = 0.02,~\xi=0.5$  &    9.6\\
$\eps = 0.02$           &    7.2\\
$\eps = 0.02,~\xi=2$    &    6.2\\
$\eps = 0.035$          &    6.2 \\
$\eps = 0.06$           &    5.3\\
\hline
\end{tabular}\\
\parbox{13cm}{
\caption{\label{table3}\small Predictions for the total cross sections in the
ZEUS untagged setup, $3 < p_T < 12$ GeV and $-1.5 < \eta < 1$.}
}
\end{center}
\end{table}

The total cross sections, obtained by integrating the curves in fig.
\ref{fig2}  over the pseudorapidity, are also shown in Table \ref{table3}. They
are to be compared with the experimental result from ZEUS \cite{zeuswwa} 
$\sigma = 10.6 \pm 1.7(stat.)\pm ^{1.6}_{1.3}(syst.)$ nb.  Notice that the 17\%
increase found going from $\eps=0.06$ to 0.035 is in good agreement with the
15\% estimated in \cite{zeuswwa} using the massive calculation.

\begin{figure}[t]
\begin{center}
\epsfig{file=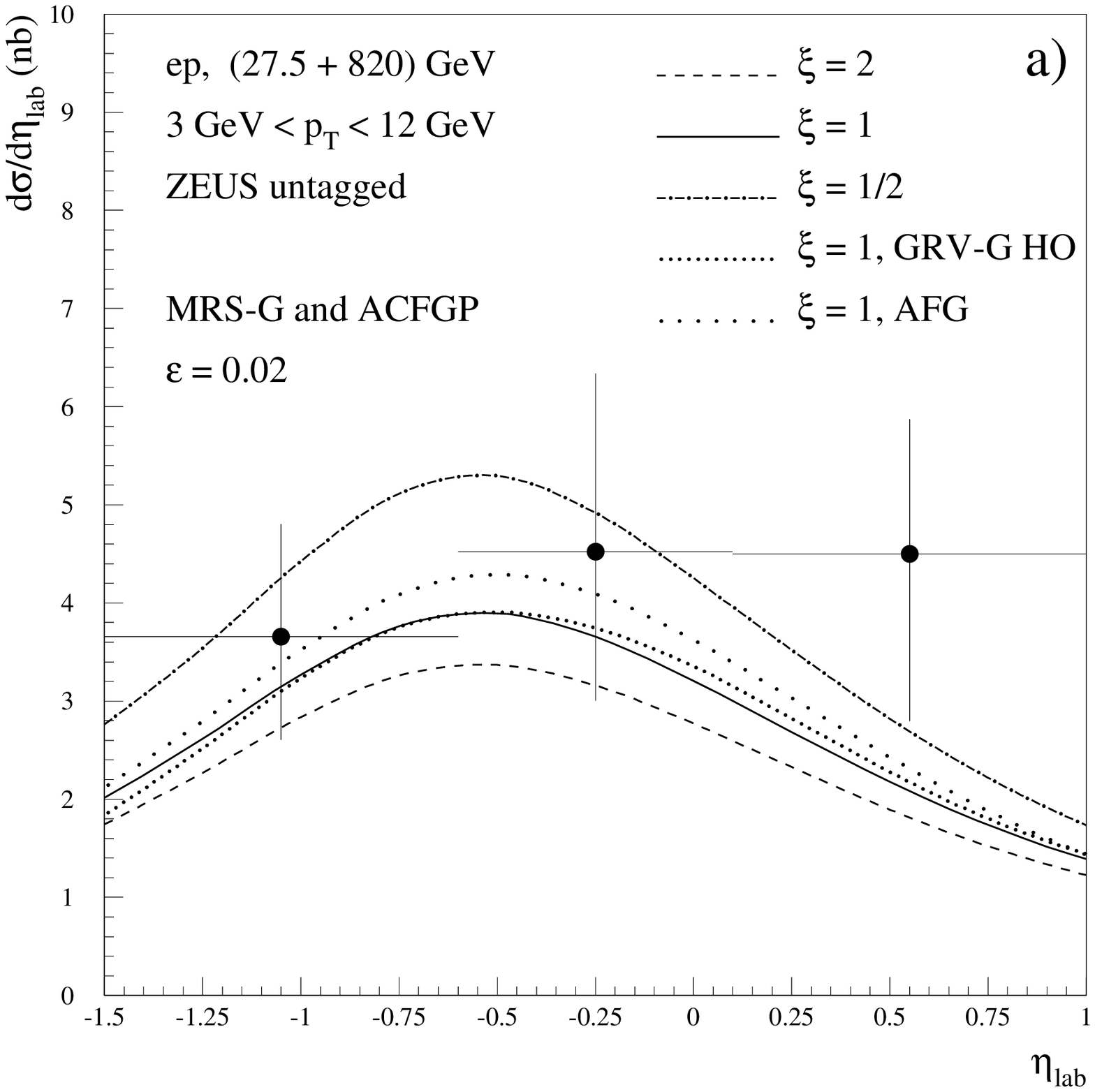,
              bbllx=30pt,bblly=160pt,bburx=540pt,bbury=660pt,
             width=10cm,clip=}
\epsfig{file=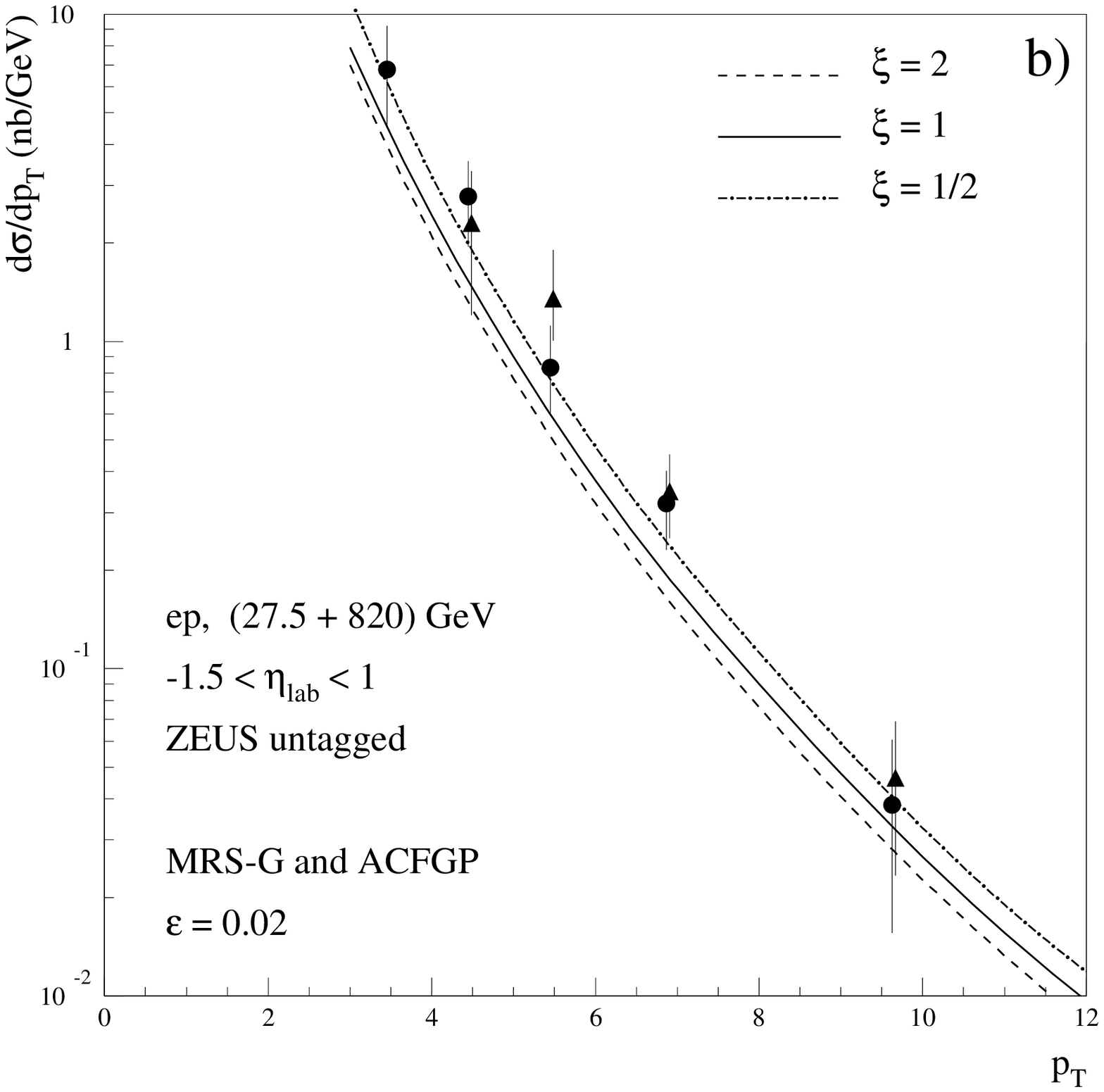,
              bbllx=30pt,bblly=160pt,bburx=540pt,bbury=660pt,
             width=10cm,clip=}
\parbox{13cm}{
\caption{\label{fig3}\small Comparison with pseudorapidity (a) and transverse
momentum (b)  experimental distributions from ZEUS \protect\cite{zeuswwa}, and
effect of variation of renormalization and factorization scales, as $\mu = \xi
m_T$, and of the photon pdf's sets.
}
}
\end{center}
\end{figure}

To get a feeling of the stability of our results we plot in fig. \ref{fig3}a
the results obtained for the pseudorapidity distribution with different choices
of renormalization/factorization scales and with the conservative value
$\eps=0.02$.  While the central curve is obtained with $\mu = m_T$, the two
others are produced with $\mu = \xi m_T$, with $\xi$ = 0.5 and 2. The variation
is not negligible, especially in the lower scale direction, but we should bear
in mind that at such a low scale we are at the border of the applicability of
perturbative QCD. Also shown on this plot are the results obtained with two
other photon pdf's sets, namely GRV-G HO and AFG \cite{afg}.  The
variations are smaller than those given by varying the scales.

By comparing with the experimental results we can see that we can get a fairly
good description of the data already with a central choice of scales.

A similar comparison is also made, in fig. \ref{fig3}b, with the $p_T$
distribution obtained by the ZEUS Collaboration. The curves, obtained 
with $\eps = 0.02$, seem to offer a fair description of the data.

\begin{figure}[t]
\begin{center}
\epsfig{file=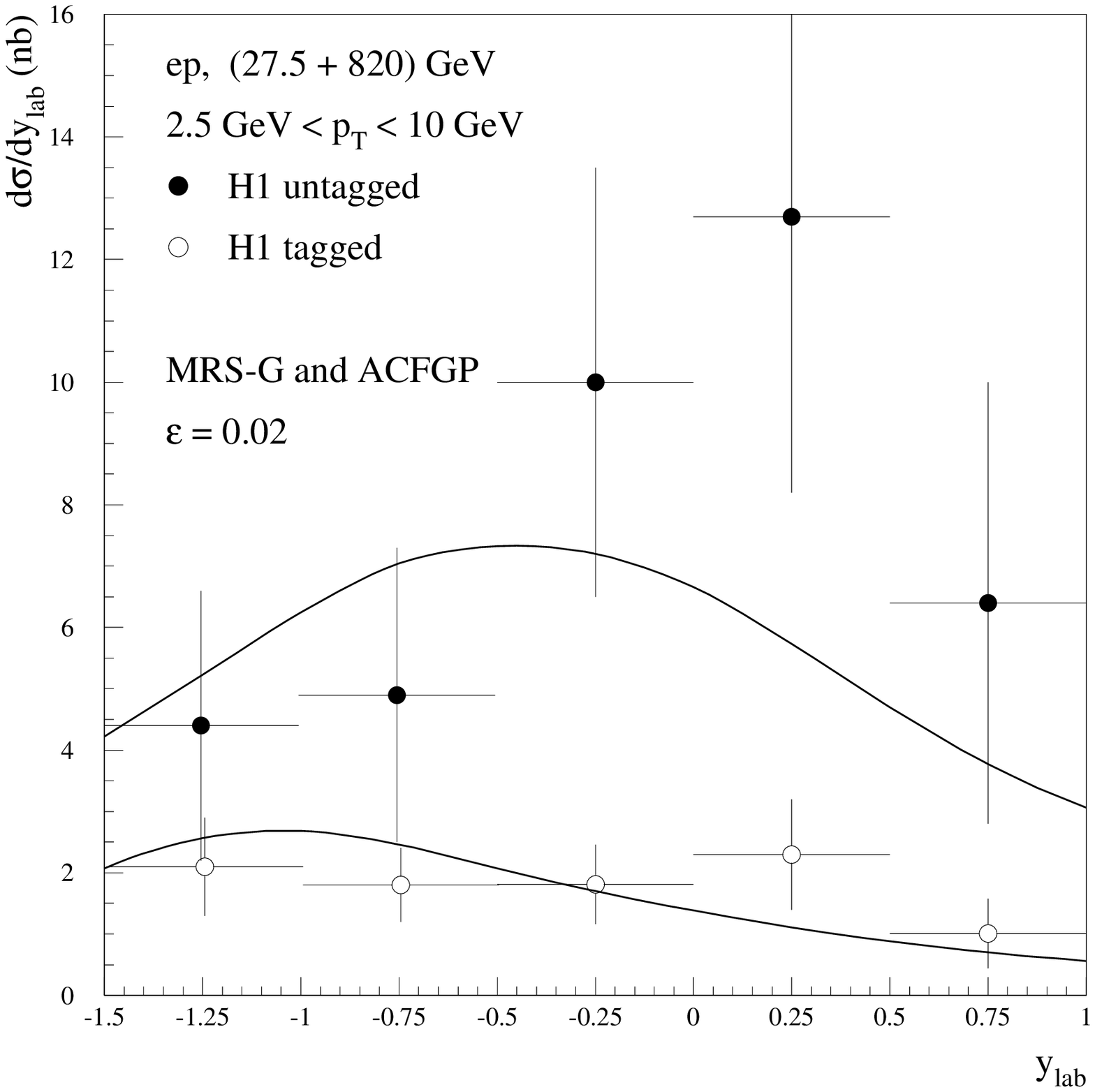,
              bbllx=30pt,bblly=160pt,bburx=540pt,bbury=660pt,
             width=10cm,clip=}
\epsfig{file=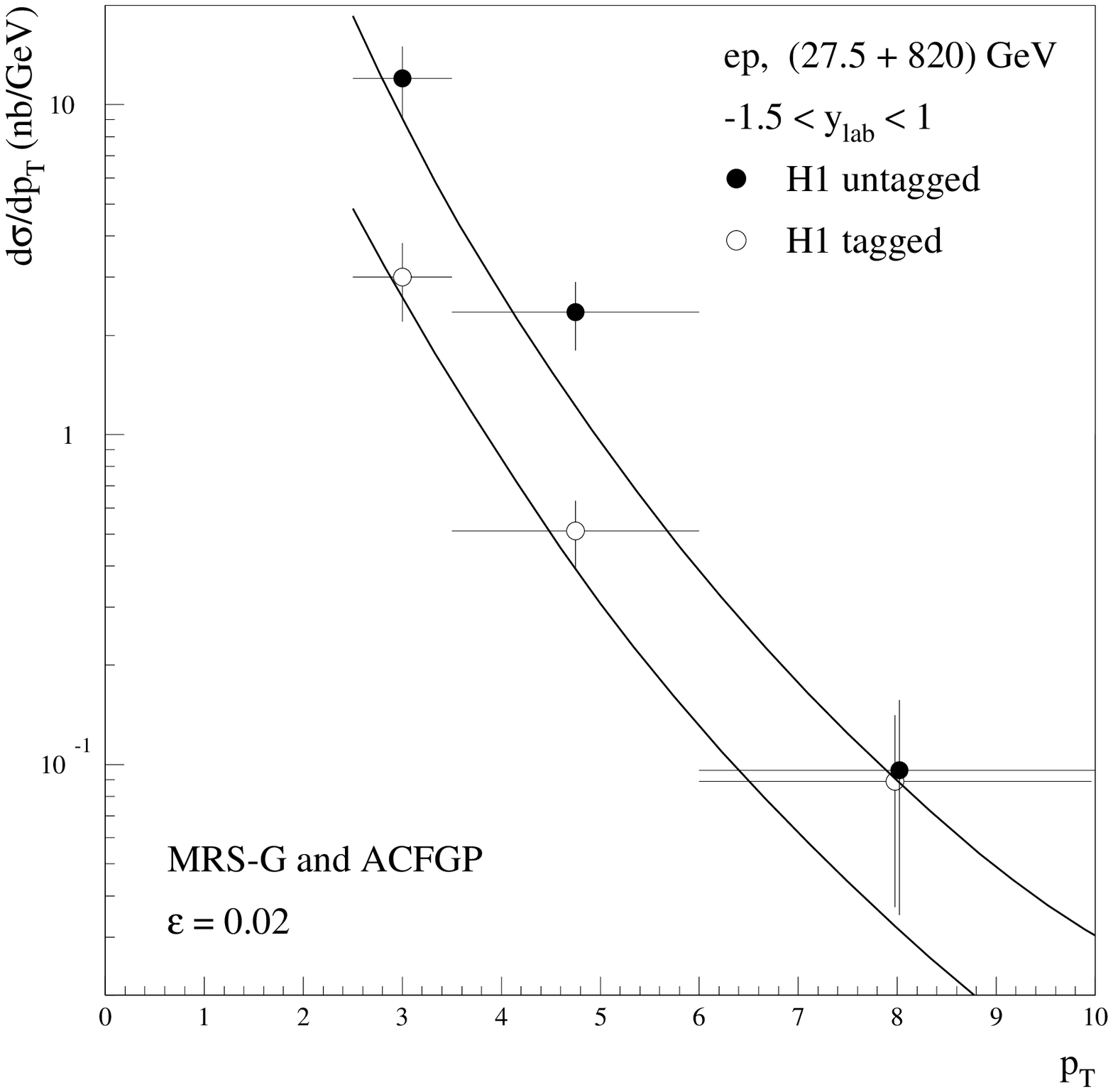,
              bbllx=30pt,bblly=160pt,bburx=540pt,bbury=660pt,
             width=10cm,clip=}
\parbox{13cm}{
\caption{\label{fig5}\small Comparison of our results with the experimental
data from the H1 Collaboration, Ref. \protect\cite{h1}.
}
}
\end{center}
\end{figure}

Finally, we want to present a comparison of the results of our approach
with more sets of data. We now use H1 results, both in the tagged and the
untagged experimental setup. These we reproduce by taking in the 
Weizs\"acker-Williams convolution $Q_{max}^2 = 0.01$ GeV$^2$, $y_{min} =
0.28$,  $y_{max} = 0.65$ and $Q_{max}^2 = 4$ GeV$^2$, $y_{min} = 0.1$, 
$y_{max} = 0.8$ respectively, according to Ref. \cite{h1}. Fig. \ref{fig5}
shows the results for the rapidity distributions\footnote{
  H1 presents its
  experimental results as a function of the rapidity rather than of the
  pseudorapidity. Our approach, in that it deals with massless partons in the
  kernel cross sections, cannot distinguish between the two. The two quantities
  become of course identical in the large $p_T$ region, and at $p_T = 2.5$ GeV
  already only differ by about 10\%.}
, obtained with the
Peterson FF with $\eps=0.02$ and the standard choice of scales, i.e. $\xi=1$.

The total cross sections for these curves, integrated within the $2.5 < p_T <
10$ GeV and $-1.5 < y < 1$ range, read 4.2 nb and 14.4 nb for the
tagged and the untagged sample respectively, to be compared with the
experimental results $4.9 \pm 0.7 \pm ^{0.74}_{0.59}$ nb and $20.2 \pm 3.3 \pm
^{4.0}_{3.6}$ nb. A quite good agreement can be seen, especially for the
tagged sample.

\section{Conclusions}
\label{concl}

In this paper we have applied the technique of fragmentation functions for
heavy mesons to $D^*$ production in $e^+e^-$ and $ep$ collisions.

These fragmentation functions are made of a perturbative part, which we evolve
with next-to-leading accuracy, and a non-perturbative one, which we fit to
$e^+e^-$ data taken by ARGUS and OPAL and subsequently use to predict
photoproduction cross sections, to be compared with data by H1 and ZEUS.

When fitting $e^+e^-$ data with a Peterson non-perturbative form we find values
for the $\eps$ parameter sensibly different from the commonly accepted value 
0.06. A central value for our fits, when using NLO evolution, is $\eps = 0.02$.
This hardens the non-perturbative fragmentation function, and increases the
photoproduction cross section, bringing it in better agreement with the data.

Our photoproduction results are found in good agreement with the NLO full
massive ones, which are reliable at the low values of $p_T$ probed by the
experiments and can be taken as a benchmark for comparisons. Convoluting them
with a Peterson with a lower $\eps$ will also increase the cross section, again
producing a better agreement with the data. Slightly less conservative choices
than those made here for the renormalization/factorization scales, the photon
parton distribution functions set, the $c\to D^*$ branching ratio and the value
of $\eps$  could easily make the agreement even better.

\vspace{.7cm}
\noindent
{\bf Acknowledgements.} We wish to thank J.Ph. Guillet and M. Fontannaz who
originally provided us with the codes for massless parton scattering, and S.
Frixione for the code for the massive calculation. Useful
conversations with G. Abbiendi, C. Coldeway, M.L. Mangano and P. Nason  
are also acknowledged.

\newpage

\appendix

\section{Appendix}

In this Appendix we show how the approaches of Refs. \cite{cgrt} and
\cite{kks} are identical at the perturbative level.

Consider a cross section for producing a heavy quark of mass $m$ at the
large scale $Q$, $\sigma(Q,m)$, given by the convolution of a coefficient
function $C(Q,\mu)$ and a perturbative fragmentation function $D(\mu,m)$. 
In the Mellin  moments space we write this as a product:
\beq
\sigma(Q,m) = C(Q,\mu) D(\mu,m),
\eeq
and $\mu$ is the factorization scale. Since $D(\mu,m)$ is the fragmentation
function evolved up to the scale $\mu$, we can write it in terms of an initial
condition at a scale $\mu_0$ as
\beq
D(\mu,m) = E(\mu,\mu_0) D(\mu_0,m) D_{np}.
\eeq
The factor $E(\mu,\mu_0)$ is the so called evolution kernel, and we have now
also included a non-perturbative term $D_{np}$, for instance the Peterson FF, 
according to eq. (\ref{ansatz}). Indeed, to think it to multiply the
perturbative  initial condition or the evolved PFF is absolutely identical,
since $D(\mu)$ is in both cases simply a product of three terms.

Putting together the two equations we have
\beq
\sigma(Q,m) = C(Q,\mu) E(\mu,\mu_0) D(\mu_0,m) D_{np}, 
\label{sigma}
\eeq
which is for instance the way we write our $e^+e^-$ cross section in Mellin
space, the one in $x$-space to be found by numerical inverse Mellin transform.

If we now consider that both the coefficient functions (see for instance Ref.
\cite{nw}) and the initial conditions of the PFF's (see eqs. (\ref{DQQ}),
(\ref{DgQ}) and (\ref{DqQ})) can be calculated as series expansions in
$\alpha_s$, like
\beq
C(Q,\mu) = 1 + \alpha_s(\mu) c(Q,\mu) \qquad \mathrm{and}\qquad 
D(\mu_0,m) = 1 + \alpha_s(\mu_0) d(\mu_0,m),
\eeq
inserting these expressions into eq.(\ref{sigma}) and rearranging it, 
up to uncontrollable $O(\alpha_s^2)$ terms we can write
\beq
\sigma(Q,m) = \Big(1 + \alpha_s(\mu) c(Q,\mu) + \alpha_s(\mu_0) d(\mu_0,m)\Big) 
         E(\mu,\mu_0) D_{np}.
\eeq
This is (with the exception of $\alpha_s(\mu_0)$ which they take $\alpha_s(\mu)$
instead) the form employed in Ref. \cite{kks} when $E(\mu,\mu_0) D_{np}$ is
considered as an ``evolved'' non-perturbative FF, and with the $d(\mu_0,m)$
functions changing the coefficient function's scheme. If one takes $\mu_0 =
\mu$ the new coefficient function will be close to the cross section for 
massive quark production, containing the logarithmic terms $\log(Q/m)$. 
Indeed, the $d(\mu_0,m)$ functions had been determined in \cite{melenason}
exactly this  way, but going the opposite way round, i.e. evaluating the full
massive $\sigma(Q,m)$, extracting from it the coefficient function $c(Q,\mu)$
in the $\overline{MS}$ scheme, and defining the remaining piece to
be the initial state condition of the fragmentation function.

\newpage

\newcommand{\zp}[3]{Z.\ Phys.\ {\bf C#1} (19#2) #3}
\newcommand{\pl}[3]{Phys.\ Lett.\ {\bf B#1} (19#2) #3}
\newcommand{\plold}[3]{Phys.\ Lett.\ {\bf #1B} (19#2) #3}
\newcommand{\np}[3]{Nucl.\ Phys.\ {\bf B#1} (19#2) #3}
\newcommand{\prd}[3]{Phys.\ Rev.\ {\bf D#1} (19#2) #3}
\newcommand{\prl}[3]{Phys.\ Rev.\ Lett.\ {\bf #1} (19#2) #3}
\newcommand{\prep}[3]{Phys.\ Rep.\ {\bf C#1} (19#2) #3}
\newcommand{\niam}[3]{Nucl.\ Instr.\ and Meth.\ {\bf #1} (19#2) #3}
\newcommand{\mpl}[3]{Mod.\ Phys.\ Lett.\ {\bf A#1} (19#2) #3}

\end{document}